\documentclass[aip,jcp,showpacs,superscriptaddress,floatfix,reprint]{revtex4-1}

\usepackage{amsmath,amssymb}
\usepackage{graphicx}
\usepackage{soul,color}

\allowdisplaybreaks

\newcommand{\lu}{\lambda_\text{u}}
\newcommand{\ls}{\lambda_\text{s}}
\newcommand{\omb}{\omega_\text{b}}

\newcommand{\kT}{k_\text{B}T}
\newcommand{\avg}[1]{\left\langle #1 \right\rangle}

\renewcommand{\Re}{\operatorname{Re}}

\begin{document}

\title{
Transition State Theory for dissipative systems without a dividing surface}

\author{F. Revuelta}
\affiliation{Grupo de Sistemas Complejos, and
  Dep.~de F\'isica y Mec\'anica,
  Escuela T\'ecnica Superior de Ingenieros Agr\'onomos,
  Universidad Polit\'ecnica de Madrid, 28040 Madrid, Spain}

\author{Thomas Bartsch}
\affiliation{Department of Mathematical Sciences, Loughborough University,
  Loughborough LE11 3TU, United Kingdom}

\author{R. M. Benito}
\affiliation{Grupo de Sistemas Complejos, and
  Dep.~de F\'isica y Mec\'anica,
  Escuela T\'ecnica Superior de Ingenieros Agr\'onomos,
  Universidad Polit\'ecnica de Madrid, 28040 Madrid, Spain}

\author{F. Borondo}
\affiliation{Departamento de Qu\'imica, and
  Instituto de Ciencias Matem\'aticas CSIC-UAM-UC3M-UCM,
  Universidad Aut\'onoma de Madrid, Cantoblanco, 28049  Madrid, Spain}


\begin{abstract}
Transition State Theory is a central cornerstone in reaction dynamics.
Its key step is the identification of a dividing surface
that is crossed only once by all reactive trajectories.
This assumption is often badly violated, especially when the reactive system
is coupled to an environment.
The calculations made in this way then overestimate the reaction rate
and the results depend critically on the choice of the dividing surface.
In this Letter, we study the phase space of a stochastically driven system
close to an energetic barrier in order to identify the geometric structure
unambiguously determining the reactive trajectories, which is then
incorporated in a simple rate formula for reactions in condensed phase
that is both independent of the dividing surface and exact.
\end{abstract}


\maketitle


Transition State Theory (TST) provides the conceptual framework for
a large part of reaction rate theory.
Originally developed to describe reactivity of small
molecules \cite{Truhlar83,Truhlar96,Miller98},
it was later extended to study a wide variety of processes in very different fields
that only have in common the existence of a transition from well-defined
``reactant'' to ``product'' states
\cite{Toller85,Eckhardt95,%
Jaffe00,Koon00,Jaffe02,Uzer02,Komatsuzaki02,Waalkens04a,Waalkens04c}.
Its great success comes from its simplicity,
since it gives a straightforward answer to the two central problems
in chemical dynamics:
the identification of the reaction mechanism and a simple
approximation to the reaction rate.

The rate limiting step in many reactions is the crossing of an energetic barrier,
the top of which forms a bottleneck
that the system must cross as the reaction takes place.
If a dividing surface (DS) is placed close to this bottleneck,
reaction rates can be readily computed from the steady-state flux through it.
The TST approximation is obtained under the assumption that the reactive
classical trajectories cross the DS only once and never return.
Very often, for example when the system is strongly coupled to a noisy
environment such as a liquid solvent, this no-recrossing assumption fails,
and any conventional DS is crossed many times by a typical trajectory.
As a result, TST calculations significantly overestimate reaction rates,
and considerable effort has been devoted to the construction of a
DS that minimizes recrossing \cite{Truhlar96}.
The problem mainly derives from the fact that this special surface
plays a double role: It defines (or separates) reactant and product regions,
and it also serves to identify the reactive trajectories.
To the latter purpose the DS is ill suited.
Despite this fundamental drawback, TST remains attractive since reactive
trajectories are simply identified as those crossing the DS.
This criterion only takes into account the instantaneous velocity at the DS,
without requiring any time consuming trajectory simulation.

Important advances in TST have recently been achieved within the approach
of modern nonlinear dynamics:
A strictly recrossing free DS can be constructed in the phase space
of reactive systems with arbitrarily many degrees of freedom
\cite{Uzer02,Waalkens04a,Waalkens04c}.
These results were generalized to systems interacting with an environment
\cite{Bartsch05b,Bartsch05c,Bartsch06a,Bartsch08,Hernandez10,Kawai10,Kawai10a}.
It has been shown that the desired exact TST can be constructed,
in the harmonic limit, by using a moving DS only crossed once by the
reactive trajectories \cite{Bartsch05b,Bartsch05c}.
Accurate results are still obtained for moderate anharmonicities \cite{Bartsch06a}
and can be improved by normal form calculations \cite{Kawai10,Kawai10a}.

In this Letter, we make a step forward by presenting an explicit
description of geometric phase space structures in an anharmonic
noisy system and an analytical scheme that relies on these structures
for the computation of \emph{exact} TST reaction rates for
arbitrary multidimensional potentials coupled to a noisy environment.
The key point is the demonstration that reactive trajectories
can be rigorously identified solely from their initial conditions,
thus avoiding the choice of a (arbitrary) DS.
This is done in terms of the stable manifold associated to a
``noisy'' Transition State (TS) trajectory jiggling in phase space.
This geometric structure encodes the relevant information about the
noise in the most economical manner and it can easily be incorporated
into a rate calculation.
Our method retains the fundamental simplicity of TST,
also providing the conceptual tools to develop new computational algorithms.
An application to the simple case of the one-dimensional quartic potential
is presented as an illustration.


The Langevin equation (LE) has been widely used to model the interaction
of a reactive system with a surrounding heat bath
\cite{Haenggi90,Pechukas76,Chandler78}.
Being a classical model, this description neglects quantum effects
such as barrier tunnelling, which can be important in the case of
light particles \cite{Bothma10}, and the interaction with excited
surfaces through conical intersections \cite{Polli10}.

The dynamics of a unit mass particle defined by coordinate $x$ moving
in a one-dimensional potential $U(x)$ is given by
\begin{equation}
    \label{Langevin}
    \ddot x = - U'(x) - \gamma \dot x + \xi_\alpha(t).
\end{equation}
Here, $\xi_\alpha(t)$ is the fluctuating force exerted by the bath,
which is connected to the damping strength, $\gamma$, by the
fluctuation--dissipation theorem
\begin{equation}
    \label{flucdis}
    \left\langle \xi_\alpha(t) \xi_\alpha(t')\right\rangle_\alpha =
        2\kT \gamma\,\delta(t-t').
\end{equation}
The model potential that we have chosen to study is
\begin{equation}
    \label{Potential1D}
    U(x) = -\frac{1}{2} \omb^2 x^2 + \frac{c_4}4 x^4,
\end{equation}
although our derivation equally applies to any other one-dimensional case.
This generality will be emphasized by using $f(t)$,
equal to $-c_4x^3$ in our case, to denote any anharmonic force.
The extension to higher dimension is straightforward
and will be presented elsewhere.

\begin{figure}
  \hfill\includegraphics[width=.4\columnwidth]{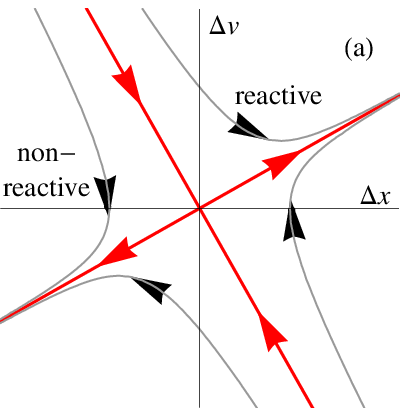}\hfill
  \includegraphics[width=.4\columnwidth]{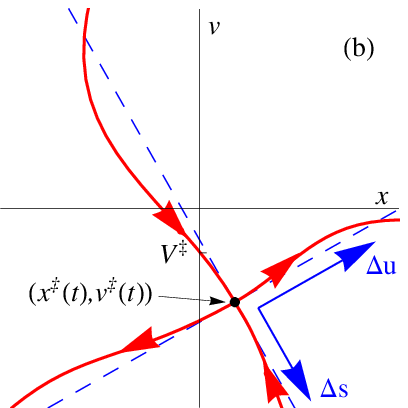}\hspace*{\fill}
  \caption{(color online)
    Schematic view of the phase space structure near the Transition State
    trajectory for the Langevin equation for the harmonic (a)
    and anharmonic (b) cases.
    The time-dependent invariant manifolds are attached to the TS~trajectory,
    and move through phase space with it.
    In the harmonic limit they appear as (red) straight lines (a),
    but they get deformed by anharmonic couplings (b).
    Reactive and non-reactive trajectories are represented in black.}
  \label{fig:mf}
\end{figure}

For every fixed realization of the noise, the LE gives rise to a specific
trajectory called TS trajectory~\cite{Bartsch05b,Bartsch05c}.
This orbit remains in the vicinity of the energetic barrier for all times,
without ever descending into any of the potential wells.
Other trajectories will be referred to the TS~trajectory,
which will be taken as a moving coordinate origin.
Since the LE is a second order differential equation,
its phase space is two-dimensional, with coordinates $x$ and $v=\dot x$.
If we now introduce the new coordinates 
\begin{align}
    \label{suTransform}
    u &= \frac{v - \ls x}{\lu-\ls}, & s &= \frac{-v+\lu x}{\lu-\ls},
\end{align}
with
\begin{equation}
  \lambda_\text{u,s} = \tfrac 12[-\gamma \pm (\gamma^2+4\omb^2)^{1/2}],
\end{equation}
relative values are defined by the time-dependent shift
\begin{equation}
    \label{relCoords}
    \Delta u = u - u^\ddag_\alpha(t), \qquad \Delta s = s - s^\ddag_\alpha(t),
\end{equation}
where the coordinates of the TS-trajectory are
\begin{align}
    \label{TStraj}
    u^\ddag_\alpha(t) &= -\frac{1}{\lu-\ls}\,\int_t^{\infty} \xi_\alpha(\tau) \,e^{\lu(t-\tau)}\, d\tau, \nonumber \\
    s^\ddag_\alpha(t) &= -\frac{1}{\lu-\ls}\,\int_{-\infty}^t \xi_\alpha(\tau)\,e^{\ls(t-\tau)}\,d\tau.
\end{align}
The corresponding equations of motion are
\begin{subequations}
\label{usDeltaEqnsOrig}
\begin{align}
    \Delta\dot u &= \lu \Delta u + \frac{f(x^\ddag_\alpha + \Delta u+\Delta s)}{\lu-\ls},
        \label{uDeltaOrig} \\
    \Delta \dot s &= \ls \Delta s - \frac{f(x^\ddag_\alpha + \Delta u+\Delta s)}{\lu-\ls},
        \label{sDeltaOrig}
\end{align}
\end{subequations}
with $x^\ddag_\alpha(t)$=$u^\ddag_\alpha(t)$+$s^\ddag_\alpha(t)$
[and $v^\ddag_\alpha(t)$=$\lu u^\ddag_\alpha(t)$+$\ls s^\ddag_\alpha(t)$].
Now, the geometric phase space structure in the vicinity of the barrier can
be easily discussed.
First, in the harmonic limit, i.e.~$f(x)=0$,
the equations of motion~\eqref{usDeltaEqnsOrig} are trivially solved, giving
\begin{align}
    \label{deltaSolHarm}
    \Delta u(t) &= \Delta u(0)\,e^{\lu t}, & \qquad
    \Delta s(t) &= \Delta s(0)\,e^{\ls t}.
\end{align}
Since $\lu>0$ and $\ls<0$, $\Delta u$ increases exponentially in time,
whereas $\Delta s$ shrinks accordingly.
More importantly, the lines $\Delta u=0$ and $\Delta s=0$ are invariant
under the dynamics of the system;
being the unstable and stable manifolds of the origin, respectively.
As shown in Fig.~\ref{fig:mf}(a), these invariant manifolds separate
trajectories with different qualitative behavior:
Those above the stable manifold (larger relative velocity)
move to the product side in the distant future,
while those below it move, on the other hand,
towards the reactant side.
%

In space fixed coordinates the invariant manifolds appear attached
to the TS~trajectory, as shown by the dashed lines in Fig.~\ref{fig:mf}(b);
their instantaneous position depends on the realization of the noise.
Accordingly, the manifolds move through phase space but
they still separate trajectories with different asymptotic behaviors.
The stable manifold intersects the $x=0$ axis in a point with
velocity $V^\ddag_\alpha$.
Trajectories with initial position $x=0$ and initial velocity $v$
larger than this critical velocity, $V^\ddag_\alpha$, are reactive,
while trajectories with initial velocities $v < V^\ddag_\alpha$ are not.
The (random) value $V^\ddag_\alpha$ therefore encodes the relevant
information about the realization of the noise concisely.
In other words, once the instantaneous position of the stable manifold is known,
any trajectory can unambiguously be classified from the values of its initial
condition as reactive or non-reactive.
Finally,
the presence of anharmonicities ($f(x) \ne 0$) will distort the invariant manifolds,
as indicated by the red lines with arrows in Fig.~\ref{fig:mf}(b).
The main step of the theory to be developed here is the calculation of this deformation.

%
%

This critical velocity can be calculated from the condition that the trajectory
with $x(0)=0$  and $v(0)=V^\ddag_\alpha$ is contained in the stable manifold
of the TS~trajectory.
To find this trajectory, which will be called \emph{critical trajectory},
we formally solve the equations of motion~\eqref{usDeltaEqnsOrig} by
\begin{subequations}
\label{intEqns}
\begin{align}
    \label{intEqU}
    \Delta u(t) = C_u\, e^{\lu t} + \frac{S[\lu,f(x^\ddag+\Delta u + \Delta s); t]}{\lu-\ls}, \\
    \label{intEqS}
    \Delta s(t) = C_s\, e^{\ls t} - \frac{S[\ls,f(x^\ddag+\Delta u + \Delta s); t]}{\lu-\ls},
\end{align}
\end{subequations}
where $C_u$ and $C_s$ are two arbitrary constants, and the integral operator
\begin{equation}
    \label{SDef}
    S_\tau[\mu, g;t] = \begin{cases}
            \displaystyle -\int_t^\infty g(\tau)\,\exp[\mu(t-\tau)] \,d\tau \!\!\!
                & :\; \Re\mu>0, \\[3ex]
            \displaystyle +\int_0^t      g(\tau)\,\exp[\mu(t-\tau)] \,d\tau \!\!\!
                & :\; \Re\mu<0.
        \end{cases}
\end{equation}
has been introduced as a convenient shorthand notation.
The subscript $\tau$ indicating the integration variable in Eq.~\eqref{SDef}
will be left out whenever this does not cause any ambiguity.
The unknown constants $C_u$ and $C_s$ can be determined by noticing that the
critical trajectory should approach the TS~trajectory for large times.
In particular, it should remain bounded for $t\to\infty$.
This can only be satisfied if $C_u=0$.
[This condition also ensures that the $S$~functional in Eq.~\eqref{intEqU} is
well defined.]
With this choice, Eq.~\eqref{intEqU} determines the initial condition $\Delta u(0)$,
and $C_s=\Delta s(0)$ can then be found from the condition that
$x(0)=x^\ddag_\alpha(0)+\Delta u(0)+\Delta s(0)=0$.
Finally, the coordinate transformation~\eqref{suTransform} yields
\begin{align}
    \label{VCritU}
    V^\ddag_\alpha = v(0) = (\lu-\ls) u(0)
\end{align}

In general, Eq.~\eqref{intEqns} represents only a formal solution,
since its right-hand side depends on the unknown functions
$\Delta u$ and $\Delta s$.
However, since it is known that the
critical trajectory remains in the neighborhood of the barrier at all times,
the anharmonic force will be small, and then its influence can be evaluated
through perturbation theory, thus obtaining an expansion
$V_\alpha^\ddag = V_{0}^\ddag + c_4 V_{1}^\ddag + c_4^2 V_{2}^\ddag + \dots$
in powers of the anharmonic coupling parameter $c_4$.

\begin{figure}
  \includegraphics[width=.8\columnwidth]{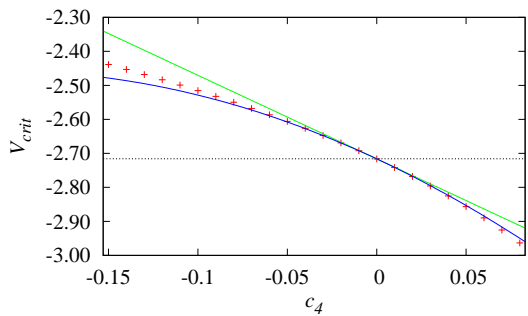}
  \caption{Critical velocity for one realization of the noise
   as a function of the anharmonicity
   for $\omb=1$, $\gamma=2.5$, and $\kT =1$:
   Numerical simulation results (red crosses), 
   harmonic approximation (gray  line),
   perturbative results to first-order (green straight line), and
   second-order (blue line).}
  \label{fig:VCritQuartic}
\end{figure}

In the harmonic approximation the critical trajectory is given by
\begin{equation}
    \label{DeltaHarm}
    \Delta u_0(t)= 0 \qquad \text{and} \qquad
    \Delta s_0(t) = -x^\ddag(0) e^{\ls t},
\end{equation}
and for this case Eq.~\eqref{VCritU} yields
\[
    V_\alpha^\ddag \equiv V^\ddag_{0}=(\lu-\ls)u^\ddag(0)
\]
that was already derived in Ref.~\onlinecite{Bartsch08}.
When the solution~\eqref{DeltaHarm} is substituted into Eq.~\eqref{intEqns},
$x=x^\ddag+\Delta u + \Delta s$ is replaced by
\begin{equation}
    \label{XDef}
    X_\alpha (t) = x_\alpha^\ddag(t)-e^{\ls t}x_\alpha^\ddag(0),
\end{equation}
which is the harmonic approximation to the coordinate $x(t)$ of the critical trajectory.
Equation \eqref{intEqU} then gives
\[
    \Delta u_1(t) = \frac{1}{\lu-\ls} S[\lu,f(X); t],
\]
and therefore the leading-order velocity correction is
\begin{align}
    \label{VCritLead}
    V^\ddag_{1}
        = S[\lu, f(X);0] = - c_4 S[\lu,X_\alpha^3;0]
\end{align}
for the quartic potential~\eqref{Potential1D}.
Given the initial condition $\Delta s_1(0)=x^\ddag_\alpha(0)-\Delta u_1(0)$,
$\Delta s_1(t)$ can be obtained from Eq.~\eqref{intEqS}, and then substituted
into Eq.~\eqref{intEqU} to find $\Delta u_2(t)$.
In this way, the second-order velocity correction
\begin{multline}
    \label{V4quartic}
    V^\ddag_2 =
     - \frac{3 c_4^2}{\lu-\ls} S_\tau\Big[\lu, X^2(\tau)  \left(
            e^{\ls \tau} S[\lu,X^3;0]  \right.\\
            \left. - S[\lu,X^3;\tau]
                + \bar S[\ls,X^3;\tau]\right); 0 \Big]
\end{multline}
is obtained.
In Fig.~\ref{fig:VCritQuartic} we present a comparison between
numerical (exact) results for the critical velocity, $V_\alpha^\ddag$,
for one realization of the noise and the approximate values computed using
\eqref{VCritLead} and~\eqref{V4quartic}, as a function of the
anharmonic parameter, $c_4$.

In order to calculate the corresponding reaction rate, $k$,
we choose the simplest DS, defined by $x=0$,
and use the basic flux-over-population rate formula (see e.g.~\cite{Haenggi90}),
which states that $k$ is proportional to the reactive flux
\begin{equation}
    \label{fluxDef}
    J = \avg{v}^\text{react}_{\alpha,v}
\end{equation}
across the DS.
This flux is to be averaged over different realizations of the noise, $\alpha$,
and also over a Boltzmann ensemble of initial velocities, $v$,
for trajectories starting at the DS.
Notice that only reactive trajectories should be included in the average.

The non-recrossing assumption of conventional TST can be restated here
by saying that the reactive trajectories are those crossing the DS
with velocity $v>0$.
We call the rate constant obtained with this approximation $k^\text{TST}$.
Any effects beyond TST are customarily summarized~\cite{Haenggi90} into a transmission
coefficient $\kappa=k/k^\text{TST}$.

\begin{figure}
 \includegraphics[width=.75\columnwidth]{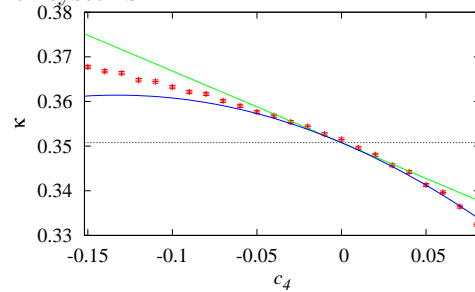}
 \caption{Transmission factor as a function of the anharmonicity
   for the same values and legend of Fig.~\protect\ref{fig:VCritQuartic}.}
\label{fig:kappaQuartic}
\end{figure}

In terms of our stochastic invariant manifolds, reactive trajectories
are characterized by $v > V^\ddag_\alpha$, as discussed before.
Using this criterion, the Boltzmann average over velocities in Eq.~\eqref{fluxDef}
can be evaluated, as it was  in Ref.~\onlinecite{Bartsch08}
for the harmonic case. This gives the exact expression
\begin{equation}
    \label{kappaVel}
    \kappa = \avg{ \exp\left(-\frac{V_\alpha^{\ddag2}}{2\kT}\right) }_\alpha,
\end{equation}
where only the average over the noise remains to be done.
By substituting  the perturbative expansion for the critical velocity
into this expression and expanding the exponential,
a perturbative series of rate corrections,
$\kappa = \kappa_0 + c_4 \kappa_1 + c_4^2 \kappa_2 + \dots$,
is obtained, where
\begin{subequations}
\label{kappaAvg}
\begin{align}
    \kappa_0 &= \avg{ E }_\alpha, \label{kappaAvg0} \\
    \kappa_1 &= -\frac{1}{\kT} \avg{ E V_0^\ddag V_1^\ddag}_\alpha, \label{kappaAvg1}
\end{align}
\end{subequations}
with the abbreviated notation
\begin{equation}
    \label{EDef}
    E = \exp\left(-\frac{V_0^{\ddag2}}{2\kT}\right)
        = \exp\left[-\frac{(\lu-\ls)^2\,u_\alpha^{\ddag2}(0)} {2\kT}\right].
\end{equation}
Expressions similar to \eqref{kappaAvg} can be obtained for the higher-order
corrections.
The leading order $\kappa_0$ was evaluated in Ref.~\onlinecite{Bartsch08}.
It yields the well-known Kramers result for the transmission factor.
With the result~\eqref{VCritLead}, the first correction term is given by
\begin{equation}
    \label{kappa2First}
    \kappa_1 = \frac{c_4(\lu-\ls)}{\kT}\,
        S_\tau \left[\lu, \avg{E\,u_\alpha^\ddag(0) X_\alpha^3(\tau)}_\alpha ; 0\right].
\end{equation}
To perform the remaining average and subsequently evaluate the $S$~functional,
note that $u^\ddag_\alpha(t)$ and $s^\ddag_\alpha(t)$,
and consequently $X_\alpha(t)$, are Gaussian random processes,
whose correlation functions were given in~\onlinecite{Bartsch05c}.
Details of this calculation are irrelevant for the purpose of this Letter
and will be presented elsewhere.
The final result is given by
\begin{equation}
    \label{kappa2Quartic}
    \kappa_1 =
     -\frac{3}{4} \frac{c_4 \, \kT}{\omb^4}\,\mu
        \left(\frac{1-\mu^2}{1+\mu^2}\right)^2
\end{equation}
in terms of the dimensionless parameter $\mu=\lu/\omb$.
This expression agrees with the corrections given in
Refs.~\onlinecite{Pollak93a,Talkner93,Talkner94a}.
A comparison between perturbation theory
and the results of a numerical simulation is shown in Fig.~\ref{fig:kappaQuartic},
along with the second-order perturbative correction,
which can be obtained in a similar manner.
The perturbative results describe the rate correctly as long as the coupling
is not too strong.
For large negative values of $c_4$ the second-order correction loses its accuracy.
By contrast, it is accurate for all positive $c_4$ shown in the figure.
If $c_4$ is increased further, the wells of the model potential~\eqref{Potential1D}
become too shallow for a rate theory to be meaningful.

In summary, we have demonstrated that the use of stochastic
invariant structures allows a rigorous TST reaction rate calculation
in general anharmonic and noisy systems.
In our approach, the DS is only used to define reactant and product regions.
To identify reactive trajectories we employ the stable manifold that is
determined by the dynamics of the system itself.
The resulting rate formula~\eqref{kappaVel} is not only remarkably compact,
but also exact.
In particular, the arbitrariness that is usually introduced by the choice
of a DS is absent.
Although our presentation is based on an analytical perturbation expansion,
the method can easily be incorporated into a numerical scheme to achieve
an efficient rate calculation in complex systems.

Support from MICINN--Spain under Contracts nrs.~MTM2009--14621 and
i--MATH CSD2006--32 is gratefully acknowledged.
FR thanks UPM for a doctoral fellowship and the hospitality of the members of
the School of Mathematics at Loughborough University, where part of this
work was done.

%

\end{document}